\documentclass[showpacs,aps,graphicx,twocolumn]{revtex4}%and
\usepackage{graphicx}

\begin{document}

\title{Controlled Teleportation of an Arbitrary Multi-Qudit State in a General Form
with $d$-Dimensional Greenberger-Horne-Zeilinger
States\footnote{Published in \emph{Chin. Phys. Lett.} \textbf{24}
(5), 1151-1153 (2007).}}
\author{Xi-Han Li$^{1,2,3}$,
 Fu-Guo Deng$^{1,2,3}$\footnote{Email address:
fgdeng@bnu.edu.cn} and Hong-Yu Zhou$^{1,2,3}$}
\address{$^1$ The Key Laboratory of Beam Technology and Material
Modification of Ministry of Education, Beijing Normal University,
Beijing 100875\\
$^2$ Institute of Low Energy Nuclear Physics, and Department of Material Science and Engineering, Beijing Normal
University,
Beijing 100875\\
$^3$ Beijing Radiation Center, Beijing 100875}
\date{\today }

\begin{abstract}
A general scheme for controlled teleportation of an arbitrary
multi-qudit state with $d$-dimensional Greenberger-Horne-Zeilinger
(GHZ) states is proposed. For an arbitrary $m$-qudit state, the
sender Alice performs $m$ generalized Bell-state projective
measurements on her 2$m$ qudits and the controllers need only take
some single-particle measurements. The receiver Charlie can
reconstruct the unknown $m$-qudit state by performing some
single-qudit unitary operations on her particles if she cooperates
with all the controllers. As the quantum channel is a sequence of
maximally entangled GHZ states, the intrinsic efficiency for qudits
in this scheme approaches 100\% in principle.
\end{abstract}

 \maketitle

Since Bennett \emph{et al.} \cite{93} presented the first protocol
in 1993, quantum teleportation, one of the most amazing features of
quantum mechanics, has been studied by many groups
\cite{guo1,zhengsb,sun,zlz,zhangzj2,deng1,yang,zhan,yanteleportation,caohj}.
The quantum teleportation process allows the two remote parties
teleport an unknown quantum state, utilizing the nonlocal
correlation of the quantum channel by means of multiparticle joint
measurement. The first controlled teleportation protocol was
presented in 1999 by using three-qubit Greenberger-Horne-Zeilinger
(GHZ) state to teleport a single-qubit state \cite{first}. In fact,
the basic idea of a controlled teleportation
 scheme \cite{zhangzj2,deng1,yang} is to let an unknown quantum state
be regenerated by a receiver with the help of controllers on a
network. This is similar to another branch of quantum communication,
called quantum state sharing (QSTS)
\cite{zhangzj1,deng2,zhangzj3,mine1,dengepjd,peng,zhangcp}, whose
task is to let several receivers share an unknown quantum state with
collaborations. Essentially one receiver reconstructs the original
state with the help of others. In general, almost all those QSTS
schemes \cite{deng2,dengepjd,mine1,zhangcp} can be used for
controlled teleportation with or without a little modification.

Recently, the controlled teleportation for a single-qubit or
$m$-qubit state has been studied. In 2004, Li \emph{et al.}
\cite{peng} proposed a scheme to share an unknown single-qubit state
with a multi-particle joint measurement. Yang \emph{et al.}
\cite{yang} presented a multiparty controlled teleportation protocol
to teleport multi-qubit quantum information. Deng \emph{et al.}
\cite{deng1} introduced a symmetric multiparty controlled
teleportation scheme for an arbitrary two-particle entanglement
state. Moreover, they presented another scheme for sharing an
arbitrary two-particle state with Einstein-Podolsky-Rosen (EPR)
pairs and GHZ state measurements \cite{deng2} or Bell-state
measurements \cite{dengepjd}. Also, Zhang, Jin and Zhang
\cite{zhangcp} presented a scheme for sharing an arbitrary
two-particle state based on entanglement swapping. Zhang \emph{et
al.} \cite{zhangzj1} proposed a multiparty QSTS scheme for sharing
an unknown single-qubit state with photon pairs and a controlled
teleportation scheme using quantum secret sharing of classical
message for teleporting arbitrary $m$-qubit quantum information
\cite{zhangzj2}. We proposed an efficient symmetric multiparty QSTS
protocol for sharing an arbitrary $m$-qubit state \cite{mine1}. More
recently, Zhan \emph{et al.} \cite{zhan} proposed a scheme to
teleport a multi-qudit cat-like state, and Zhang \emph{et
al.}\cite{zhangzj3} proposed a QSTS protocol for sharing an unknown
single-particle qutrit state (the state of a three-level quantum
system).

In this Letter, we give a general scheme for controlled
teleportation of an arbitrary $m$-qudit state with $d$-dimensional
GHZ states via  the control of $n$ controllers, following some ideas
in Ref. \cite{mine1}. Except for the sender Alice, each of the
controllers needs only to take $m$ single-particle measurements on
his particles, and the receiver can reconstruct the unknown
$m$-qudit state with $m$ unitary operations if she cooperates with
all the controllers. This scheme for controlled teleportation of
$m$-qudit state is optimal as its intrinsic efficiency for qudits
$\eta_q\equiv \frac{q_u}{q_t}$ approaches 100\% in principle. Here
$q_u$ is the number of the useful qubits and $q_t$ is the number of
the qubits transmitted.

For a $d$-dimensional Hilbert space, its basis along z-direction
$Z_d$, which has $d$ eigenvectors, can be written as
\begin{eqnarray}
\vert 0 \rangle,\;\;\;\; \vert 1 \rangle, \;\;\;\;\vert 2
\rangle,\;\;\;\;\cdots, \;\;\;\; \vert d-1 \rangle
\end{eqnarray}
The $d$ eigenvectors of another unbiased measuring basis (MB) $X_d$
can be described as \cite{licy1,lixhjkps}
\begin{eqnarray}
\vert u \rangle_x
=\frac{1}{\sqrt{d}}\sum\limits_{l=0}^{d-1}{e^{\frac{2\pi i}{d}
ul}\vert l \rangle},
\end{eqnarray}
where $u=0,1,\dots,d-1$,  i.e.,
\begin{eqnarray}
\vert 0\rangle_x&=&\frac{1}{{\sqrt d }}\left( {\left\vert  0
\right\rangle + \vert 1\rangle \;\; + \cdots \;\; + \left\vert
{d-1}\right\rangle }\right),\;\nonumber \\
\vert 1\rangle_x&=&\frac{1}{{\sqrt d }}\left({\left\vert  0
\right\rangle + e^{{\textstyle{{2\pi i} \over d}}} \left\vert  1
\right\rangle + \cdots
+ e^{{\textstyle{{(d-1)2\pi i} \over d}}} \left\vert  {d-1} \right\rangle} \right),\; \nonumber\\
&&\cdots \cdots \cdots \cdots \cdots \cdots \nonumber\\
\vert d-1\rangle_x&=&\frac{1}{{\sqrt d }}(\left\vert  0
\right\rangle + e ^{{\textstyle{{2(d-1)\pi i} \over d}}} \left\vert
1 \right\rangle  + e ^{{\textstyle{{2\times 2(d-1)\pi i} \over d}}}
\left\vert 2 \right\rangle \nonumber\\
&& + \cdots + e^{{\textstyle{{(d-1)\times 2(d-1)\pi i} \over d}}}
\left\vert {d-1} \right\rangle ).
\end{eqnarray}
The two vectors $\vert k\rangle$ and $\vert l\rangle_x$ coming from
two MBs satisfy the relation $\vert \langle k|l\rangle_x \vert
^2=\frac{1}{d}$.  The generalized $d$-dimensional Bell states are
\cite{wangc,liuxspra}
\begin{eqnarray}
\vert \psi_{uv} \rangle=\frac{1}{\sqrt{d}}
\sum\limits_{l=0}^{d-1}{e^{\frac{2\pi i}{d} lu}\vert l \rangle
\otimes \vert l \oplus v \rangle },
\end{eqnarray}
where  $u,v=0,1,...,d-1$. The $d$-dimensional unitary operation
\begin{eqnarray}
U_{uv}=\sum\limits_{l=0}^{d-1}{e^{\frac{2\pi i}{d} ul}\vert l \oplus
v \rangle \langle l|}
\end{eqnarray}
can transform the generalized $d$-dimensional Bell state (GBS)
\begin{eqnarray}
\vert \psi_{00} \rangle=\frac{1}{\sqrt{d}}
\sum\limits_{l=0}^{d-1}{\vert l \rangle \otimes \vert l \rangle }
\end{eqnarray}
into the GBS $\vert \psi_{uv} \rangle$, i.e.,
\begin{eqnarray}
U_{uv}\vert \psi_{00} \rangle=\vert \psi_{uv} \rangle.
\end{eqnarray}

\begin{figure}[!h]%[tpb]
\begin{center}
\includegraphics[width=8cm,angle=0]{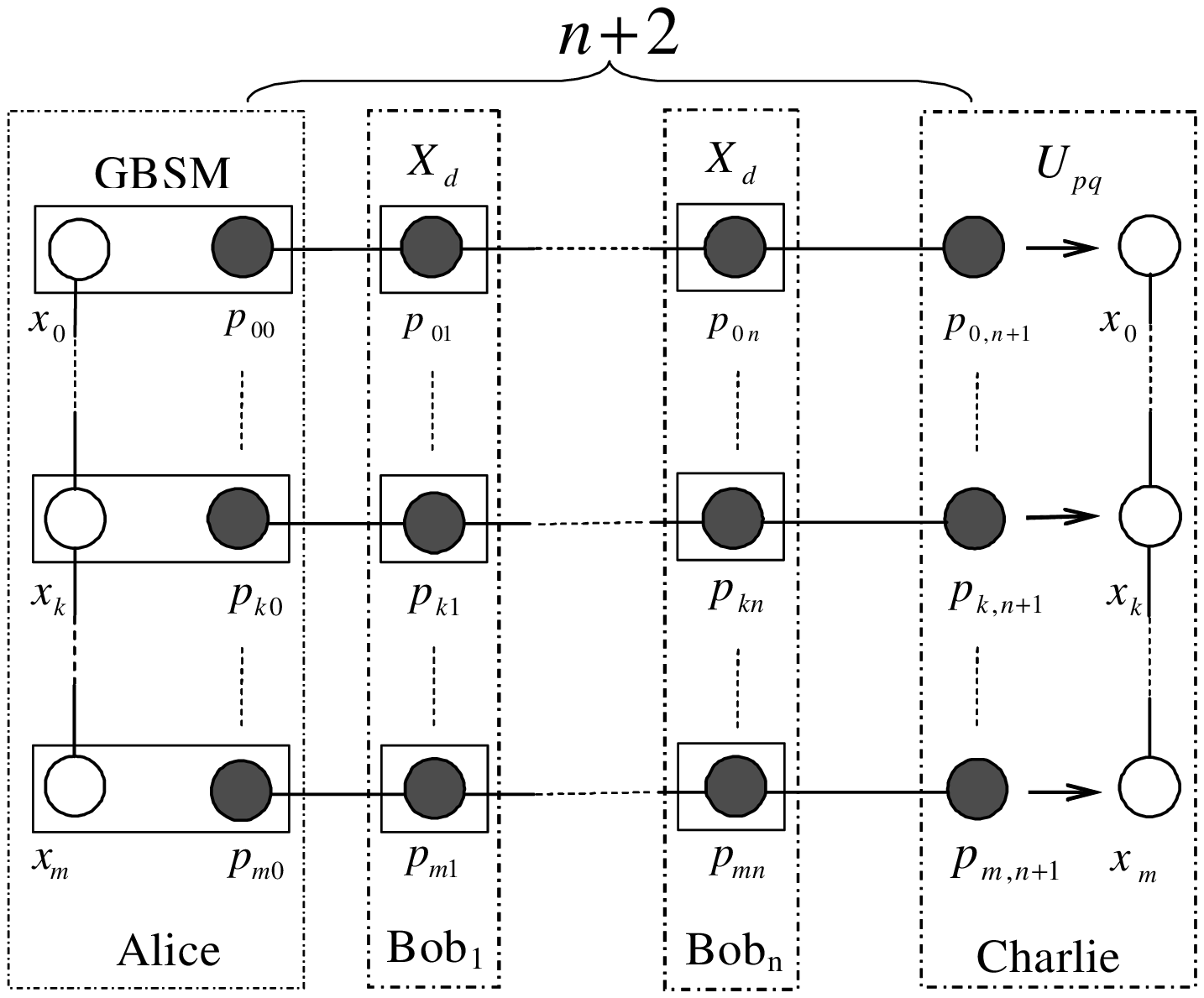} \label{f1}
\caption{Schematic demonstration of controlled teleportation of an
arbitrary $m$-qudit state by using $m$ ($n+2$)-particle GHZ states
as the quantum channel.  }
\end{center}
\end{figure}

Now, let us describe the  process of our controlled teleportation in
detail. An unknown arbitrary $m$-qudit state teleported can be
written as
\begin{eqnarray}
\vert \phi \rangle_{x_1,x_2,...,x_m} = \sum\limits_{x_1,x_2,...,x_m
= 0}^{d - 1} {a_{x_1,x_2,...,x_m} \vert x_1, x_2,...,x_m \rangle}.
\end{eqnarray}
The complex coefficients $a_{x_1,x_2,...,x_m}(x_{k}=0,1,...,d-1)$
are unknown and satisfy the  normalized relation
\begin{eqnarray}
\sum\limits_{x_1,x_2,...,x_m = 0}^{d -
1}{|a_{x_1,x_2,...,x_m}|^2=1}.
\end{eqnarray}
For teleporting the unknown $m$-qudit state $\vert \phi
\rangle_{x_1,x_2,...,x_m}$, the sender Alice, the receiver Charlie
and the $n$ controllers (called Bob$_j$, $j=1,2,...,n$) should share
securely $m$ $(n+2)$-particle $d$-dimensional GHZ states in advance,
see Fig.1. The $k$-th ($k=1,2,...,m$) GHZ state is
\begin{eqnarray}
\vert \Psi_{GHZ} \rangle_k= \frac{1}{\sqrt{d}}
\sum\limits_{l=0}^{d-1} {\vert l,
l,...,l\rangle}_{p_{k0},p_{k1},\dots, p_{k,n+1}}.
\end{eqnarray}
For each GHZ state $\vert \Psi_{GHZ} \rangle_k$, the particles
$p_{k0}$ ($p_{k,n+1}$) are held in the hands of Alice (Charlie), and
the particles $p_{kj}$ are controlled by the controller Bob$_j
(j=1,2,...,n)$. The composite state of the quantum system composed
of the particles in an unknown state $\vert \phi
\rangle_{x_1,x_2,...,x_m}$ and those in all the GHZ states is
\begin{eqnarray}
\vert \Psi_{all} \rangle=\vert \phi \rangle_{x_1,x_2,...,x_m}
\otimes \prod\limits_{k= 1}^m \vert \Psi_{GHZ} \rangle_k.
\end{eqnarray}

The sender Alice first performs $m$ generalized Bell-state
measurements (GBSM) on the particles $x_k$ and $p_{k0}$
$(k=1,2,...,m)$. Suppose the results of the $m$ GSBM are $\vert
\psi_{\alpha_{k1}\alpha_{k2}}\rangle$ ($k=1,2,...,m$), and then the
state of the quantum system composed of the  particles controlled by
Bob$_{j}$ ($j=1,2,\ldots, n$) and Charlie becomes (without being
normalized)
\begin{eqnarray}
\nonumber \vert \Psi'_{sub} \rangle_{BC}=\frac{1}{d^m}
\sum\limits_{l_1,l_2,...,l_m=0}^{d-1}{e^{-\frac{2\pi i}{d}
\sum\limits_{k=1}^{m}l_k\alpha_{k1} }} a_{l_1,l_2,...,l_m}\\
\prod_{k=1}^{m}\vert \underbrace{ l_k+ \alpha_{k2},...,l_k+ \alpha_{k2}}_{n+1}\rangle_{p_{k1}, p_{k2},..., p_{k,n+1}}.%\nonumber\\
\end{eqnarray}
When Charlie wants to reconstruct the original state, he asks
Bob$_j$ to measure the particles $p_{kj}(k=0,1,...m)$ with the $X_d$
basis independently. If the result of the  measurement on the
particle $p_{kj}$ is $\vert \beta_{kj}\rangle_x$, the retained
subsystem composed of the particles controlled by Charlie is in the
state
\begin{eqnarray}
\vert \Psi''_{sub} \rangle_{C}&=&\frac{1}{d^{m(1+\frac{n}{2})}}\nonumber\\
&&\sum\limits_{l_1,l_2,...,l_m=0}^{d-1} {e^{-\frac{2\pi i}{d}
\sum\limits_{k=1}^{m}[l_k\alpha_{k1}+(l_k+\alpha_{k2})\sum\limits_{j=1}^{n}\beta_{kj}]}}\nonumber\\
&&a_{l_1,l_2,...,l_m}  \prod\limits_{k = 1}^m\vert l_k+\alpha_{k2}
\rangle_{p_{k,n+1}}.
\end{eqnarray}
Charlie can perform a unitary operation $U_{p_{k}q_{k}}$ on each of
her particle $p_{k,n+1}$ if he wants to regenerate the unknown state
$\vert \phi \rangle_{x_1,x_2,...,x_m}$. The operations
$U_{p_{k}q_{k}}$ can be chosen according to the information about
the results of the measurements done by the sender Alice and the
controllers Bobs, i.e.,
\begin{eqnarray}
p_k=\alpha_{k1}+\sum\limits_{j=1}^{n}{\beta_{kj}}, \\
q_k=d-\alpha_{k2}.
\end{eqnarray}
After $m$ unitary operations, the state of the $m$ particles in
Charlie's hands is transformed into
\begin{eqnarray}
\vert \Psi''' \rangle_C=A\sum\limits_{l_1,l_2,...,l_m=0}^{d-1}
a_{l_1,l_2,...,l_m}\vert l_1,l_2,...,l_m\rangle,
\end{eqnarray}
where
\begin{eqnarray}
A=\frac{1}{d^{m(1+\frac{n}{2})}}{e^{\frac{2\pi
i}{d}\sum\limits_{k=1}^{m}\alpha_{k1}\alpha_{k2} }}.
\end{eqnarray}
$A$ is a whole phase and it does not affect the feature of the
state. In this way, the receiver Charlie obtains the original state
$\vert \phi \rangle_{x_1,x_2,...,x_m}$. As the quantum channel is a
sequence of maximally entangled GHZ states, all of the qudits
transmitted are useful in principle, and the intrinsic efficiency
for qudits $\eta_q$ in this scheme approaches 100\%.

To prevent some controllers from stealing the information about the
unknown state  $\vert \phi \rangle_{x_1,x_2,...,x_m}$ in this
scheme, the parties can exploit the way in Ref.
\cite{deng1,licy,licy1,dengcpl,dengcpl2,lixhcpl} to set up the
quantum channel securely. That is, Alice prepares some decoy photons
who are randomly in the states $\{\vert 0\rangle, \vert
1\rangle,\cdots,\vert d-1\rangle,\vert 0\rangle_x, \vert
1\rangle_x,\cdots,\vert d-1\rangle_x\}$, and inserts them into the
particle sequences $S_j=[p_{0j},p_{1j},\cdots,p_{mj}]$
($j=1,2,\cdots,n+1$) which are sent to Bob$_{j}$ ($j\leq n$) and
Charlie ($j= n+1$), respectively. After confirming the receipt of
the sequences, Alice picks up the decoy photons to check the
eavesdropping in the quantum line, similar to Ref.
\cite{two-step,improving,QOTP}. In detail, Alice tells the
controllers Bob$_j$ and Charlie to measure their decoy photons with
the MB $Z_d$ or the MB $X_d$, the same as those chosen by Alice for
preparing them. Thus the process for setting up the quantum channel
can be made to be secure.

In summary, we have presented  a general scheme for multiparty
controlled teleportation of an arbitrary multi-qudit state. The
quantum channel used is a sequence of $d$-dimensional GHZ states.
For teleporting an arbitrary $m$-qudit state, the sender Alice
should perform $m$ generalized Bell-state measurements, and each
controller needs to perform $m$ single-qudit measurements using the
$X_d$ basis. According to the results of the measurements published
by Alice and Bobs, the receiver Charlie can choose $m$ suitable
single-qudit unitary operations to reconstruct the original state.
As almost all the GHZ states can be used to teleport the unknown
state in principle, the intrinsic efficiency for qudits in this
scheme approaches 100\%.

This work was supported by the National Natural Science Foundation
of China under Grant Nos. 10604008 and 10435020, and Beijing
Education Committee under Grant No. XK100270454.


\begin{thebibliography}{99}
\bibitem{93}Bennett C H et al 1993 \emph{Phys. Rev. Lett.}
\textbf{70} 1895

\bibitem{guo1} Zheng S B and Guo G C 1997 \emph{Phys. Lett.} A \textbf{232} 171

\bibitem{zhengsb} Zheng S B 2006 \emph{Chin. Phys. Lett.} \textbf{23}  2356

\bibitem{sun} Yu S X and Sun C P 2000 \emph{Phys. Rev.} A \textbf{61}  022310

\bibitem{zlz} Zhan X G et al 2006 \emph{Chin. Phys. Lett.} \textbf{23}  2900


\bibitem{yang}Yang C P et al 2004 \emph{Phys. Rev.} A \textbf{70} 022329

\bibitem{deng1}Deng F G et al 2005 \emph{Phys. Rev.} A \textbf{72} 022338

\bibitem{zhangzj2}Zhang Z J 2006 \emph{Phys. Lett.} A \textbf{352} 55




\bibitem{zhan} Zhan X G et al 2006 \emph{Chin. Phys. Lett.} \textbf{23} 2900

\bibitem{yanteleportation} Yan F L and Ding H W 2006 \emph{Chin. Phys. Lett.} \textbf{23} 17

\bibitem{caohj} Cao H J et al 2006 \emph{Chin. Phys.} \textbf{15}  915



\bibitem{first}Karlsson A and Bourennane M 1998 \emph{Phys. Rev.} A \textbf{58}
4394

\bibitem{deng2}Deng F G et al 2005 \emph{Phys. Rev.} A \textbf{72} 044301

\bibitem{mine1}Li X H et al 2005 \emph{J. Phys.} B \textbf{39} 1975

\bibitem{dengepjd}Deng F G et al 2006 \emph{Eur. Phys. J.} D
\textbf{39} 459

\bibitem{zhangcp} Zhang Y Q et al 2006 \emph{Chin. Phys.} \textbf{15} 2252


\bibitem{zhangzj1}Zhang Z J 2005 \emph{Eur. Phys. J.} D \textbf{33} 133


\bibitem{zhangzj3}Wang Z Y and Zhang Z J  quant-ph/0607187


\bibitem{peng}Li Y M et al 2004 \emph{Phys. Lett.} A \textbf{324} 420



\bibitem{licy1} Li C Y et al 2006 \emph{Chin. Phys. Lett.} \textbf{23} 2896


\bibitem{lixhjkps} Li X H et al 2006 \emph{J. Korean Phys. Soc.} \textbf{49}
1354

\bibitem{wangc} Wang C et al 2005
\emph{Phys. Rev.} A \textbf{71}  044305

\bibitem{liuxspra}
Liu X S et al 2002 \emph{Phys. Rev.} A 65 022304


\bibitem{licy} Li C Y et al 2005 \emph{Chin. Phys. Lett.} \textbf{22} 1049

\bibitem{dengcpl} Deng F G et al 2006 \emph{Chin. Phys. Lett.} \textbf{23} 1676

\bibitem{dengcpl2} Deng F G et al 2006 \emph{Chin. Phys. Lett.} \textbf{23} 1084

\bibitem{lixhcpl} Li X H et al 2007 \emph{Chin. Phys. Lett.} \textbf{24} 23

\bibitem{two-step} Deng F G, Long G L and Liu X S 2003 \emph{Phys. Rev.} A
\textbf{68} 042317


\bibitem{improving} Li X H et al 2006 \emph{Phys. Rev.} A \textbf{74} 054302

\bibitem{QOTP} Deng F G and Long G L 2004 \emph{Phys. Rev.} A \textbf{69}
052319


\end{thebibliography}
\end{document}